\documentclass[conference]{IEEEtran}
\IEEEoverridecommandlockouts
\usepackage{cite}
\usepackage{amsmath,amssymb,amsfonts}
\usepackage{algorithmic}
\usepackage{graphicx}
\usepackage{textcomp}
\usepackage{xcolor}
\usepackage{ifthen}
\def\BibTeX{{\rm B\kern-.05em{\sc i\kern-.025em b}\kern-.08em
    T\kern-.1667em\lower.7ex\hbox{E}\kern-.125emX}}
    
\usepackage{pifont}
\usepackage{multirow}
    
\begin{document}

\title{Centralized vs. Decentralized Security \\ for Space AI Systems? A New Look}

\author{}

\author{\IEEEauthorblockN{Noam Schmitt}
\IEEEauthorblockA{\textit{Orange Research, IP Paris, Télécom SudParis, ENS Paris-Saclay} \\
noam.schmitt@orange.com}
\and
\IEEEauthorblockN{Marc Lacoste}
\IEEEauthorblockA{\textit{Orange Research} \\
marc.lacoste@orange.com}}

\maketitle

\begin{abstract}
This paper investigates the trade-off between centralized and decentralized security management in constellations of satellites to balance security and performance. We highlight three key AI architectures for automated security management: (a) centralized, (b) distributed and (c) federated. 
The centralized architecture is the best option short term, providing fast training, despite the hard challenge of the communication latency overhead across space. Decentralized architectures are better alternatives in the longer term, providing enhanced scalability and security.
\end{abstract}


\section{Security of Satellite Systems}

\noindent New Space is a new frontier where no AI nor Cloud had much gone before... At least until very recently. Among many: a frenzy of initiatives on future large-scale infrastructures and non-terrestrial networks (e.g., launching multiple satellite constellations, space cloud computing) from multiple competing stakeholders; stringent constraints on communications; 
multiple new threats and significant momentum to develop automated security solutions using AI. 
All those elements crystallize into an emerging new field: \textit{Space AI Security}, to detect, protect and respond to threats~\cite{europeanunionagencyforcybersecurity.LEOSatcomCyber2024::short}.

\vspace*{0.05cm}

Yet, a fundamental question remains: \textbf{which is the ``right'' AI architecture to manage security in such space systems: centralized or partly/fully decentralized?} 
With which trade-offs for security, performance (latency, scale) and utility? 

\vspace*{0.05cm}

The simplest design is based on a \textit{large central model}. 
This choice was deemed not practical due to the limited contact time between satellite and ground station, low bandwidth and high latency without inter-satellite links (ISL). 
Decentralized design alternatives around distributed variants  of \textit{Federated Learning} (FL) 
were then investigated as more viable to perform training or inference in space and viewed as more scalable and secure
~\cite{linFedSNFederatedLearning2025::short}.
Recently, advances in satellite technologies such as ISL communications -- and increasingly bandwidth -- have opened new possibilities. 
Centralized architectures seem to be making a comeback, and are now adopted by many major industry constellation systems providers. What then of the centralized vs. decentralized alternative?

In this paper, we take a closer look  
and distinguish 3 key  AI architectures for automated security management: (a) centralized, (b) distributed, and (c) federated, highlighting benefits and limitations. We then explore on a case study the centralized vs. decentralized security architecture trade-off  regarding accuracy and latency. 

\vspace*{0.2cm}

\begin{figure}[htbp]
\vspace*{-0.4cm}
\centering
\includegraphics[width=0.5\textwidth]{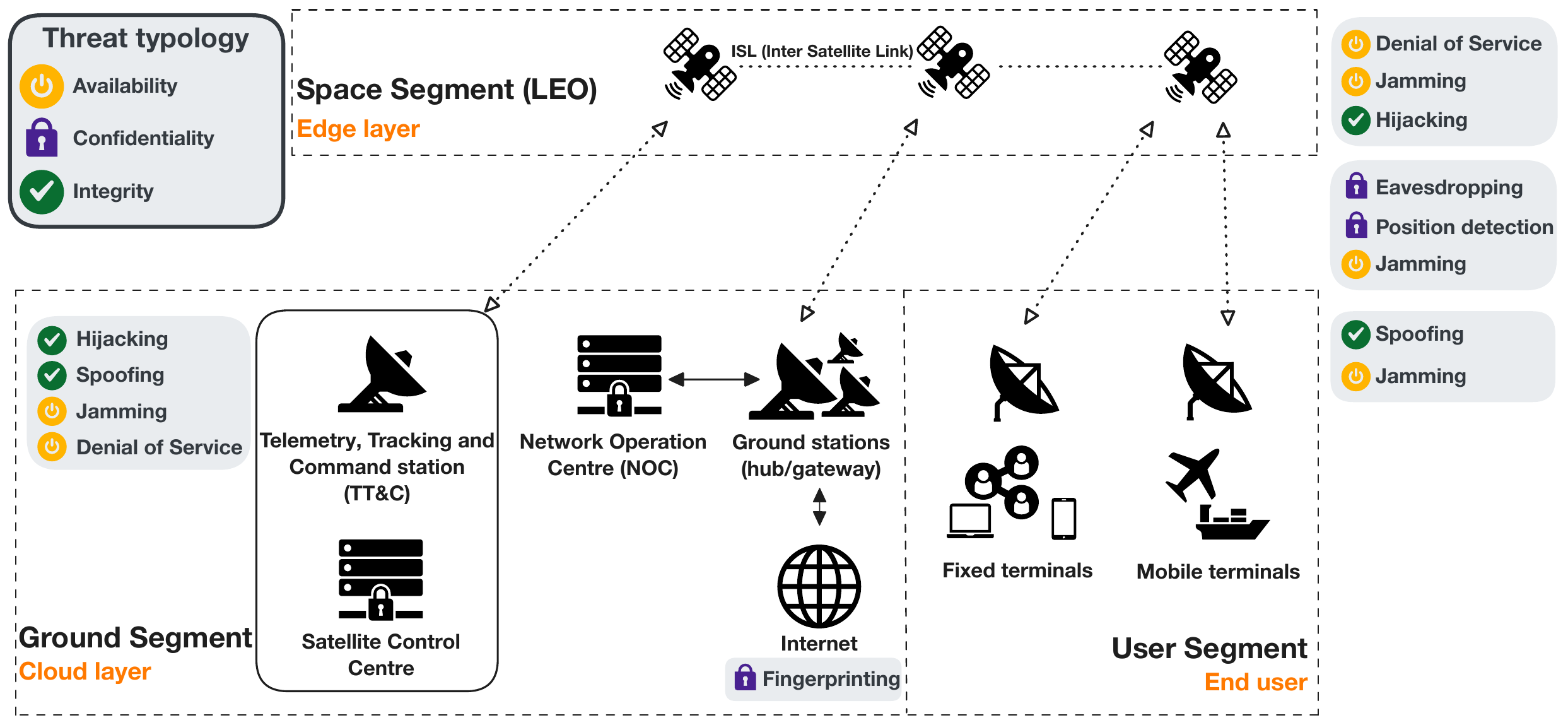}
\vspace*{-0.75cm}
\caption{Satellite constellation: architecture and main threats}
\label{fig:SpaceConstellationOverview}
\vspace*{-0.6cm}
\end{figure}

\noindent\textbf{Architecture.} Figure~\ref{fig:SpaceConstellationOverview} gives an overview of the main entities in a satellite constellation.
The \textbf{User} segment includes the end users of the service or system \textit{tenants}, deployed on fixed and mobile terminals with different risks. 
The \textbf{Space} segment includes all satellites in orbit, connected via links to the User segment, the Ground segment and neighboring satellites. They are typically deployed in a Low Earth Orbit (LEO). Some operators (e.g., OneWeb) use higher altitude orbits to make the constellation more dynamic depending on connection and load.
The \textbf{Ground} segment includes \textit{Ground stations (GS)} distributed across the planet 
and connected to the Internet to manage the satellite payload. 
The constellation is supervised by the Satellite Control Center, which notably manages orbits.

\vspace*{0.1cm}

\noindent\textbf{Threats.} 
\textit{Confidentiality} threats include 
\textit{position detection} or \textit{user tracing} specific to the satellite network architecture. \textit{Eavesdropping} deals with risks of interception and spying on communications.  \textit{Integrity} threats include \textit{spoofing}, e.g., to  usurp a signal, a location or an end-point and \textit{hijacking}.   
The main \textit{availability} threat remains \textit{Denial-of-service}, in forms common to virtualized infrastructures or specific to LEO constellations. 
Mitigations include anomaly detection, packet filtering and firewalling. \textit{Jamming}, in base stations or satellites, is another potent threat to disrupt communications. 
To prevent, detect, and mitigate those threats, a wide range of counter-measures are available~\cite{europeanunionagencyforcybersecurity.LEOSatcomCyber2024::short}.

\section{Space AI Security Architectures}

\begin{figure}[htbp]
\centering
\vspace*{-0.2cm}
\includegraphics[width=0.5\textwidth]{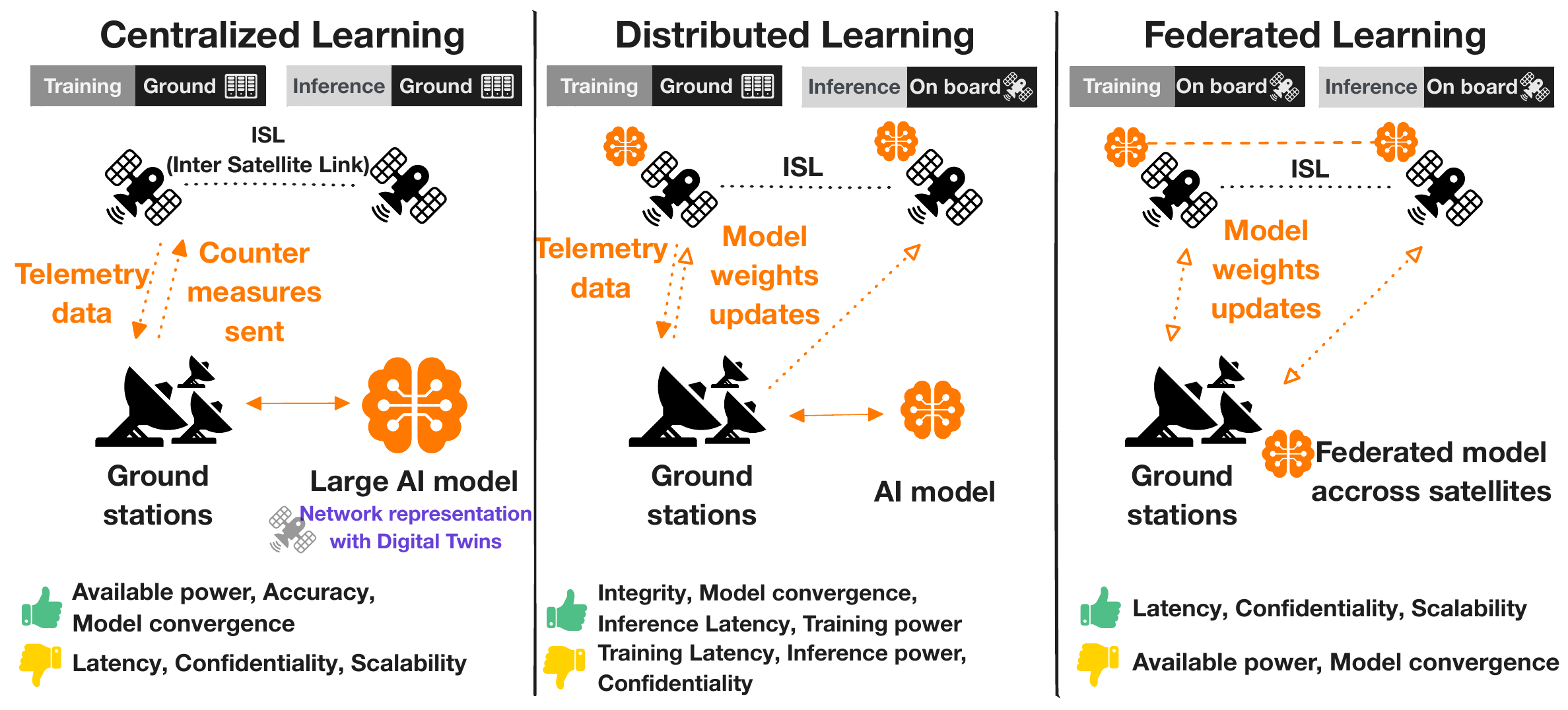}
\vspace*{-0.8cm}
\caption{Space AI architectures: (a) centralized, (b) distributed, (c) federated}
\label{fig:AIArchitectures}
\vspace*{-1.0cm}
\end{figure}

\noindent 
Satellites face stringent embedded constraints (e.g., response time, hardware architecture, energy efficiency).
For risk mitigation,
deployment of counter-measures  
must be automated and orchestrated dynamically. 
Three options are available for a distributed AI security architecture 
(see Figure~\ref{fig:AIArchitectures}).

\newpage
\vspace*{0.1cm}

\noindent\textbf{(a) Centralized Architecture.} A central AI model is deployed on the ground for training and inference. The model learns from satellite telemetric data sent from space. The ground system behaves as a digital twin of the constellation~\cite{ferreiraReinforcementLearningSatellite2019::short}. 
It triggers the required counter-measure updates, then sent back.

Deploying large models results in fast training and inference: unbounded computing and data processing are available. Additionally, there are no embedded barriers, e.g., for energy efficiency nor need for additional AI hardware on-board.

Threat response remains slow due to large latencies across space between telemetric data sending and security update. The attack surface is wide 
and scalability not guaranteed.

\vspace*{0.1cm}

\noindent\textbf{(b) Distributed Architecture.}
Satellites send telemetric data to the GS for model training as in (a), but receive weights for on-board model update. Inference is performed on-board.

On-board inference lowers latency of the security response.
The system is more protected than in design (a) against integrity threats, although confidentiality threats remain.
Embedded constraints may nonetheless lower model accuracy. 
Scalability may also remain difficult at training time.

\vspace*{0.1cm}

\noindent\textbf{(c) Federated Architecture.} 
Training and inference are performed on board, without sending telemetric data~\cite{linFedSNFederatedLearning2025::short}.
Gradients are shared with other satellites and GS for aggregation. 

This class of architecture has the lowest latency. 
It is more secure than centralized designs (a) and (b).
It should also be highly scalable up and out, notably, being embedded by design.
AI accuracy might be a challenge due to smaller models operating under embedded hardware and energy constraints.
Model convergence and satellite synchronization might also be issues for large constellations.
Federated learning specific threats should also be addressed.

\section{Experimental Results}

\noindent We compare centralized vs. decentralized designs, in terms of \textit{accuracy} and \textit{performance} during \textit{training} 
and \textit{inference}. 
All experiments are run in Python on the CIFAR-10 dataset and ResNet20 model. FL simulations with IID partitioning use the Flower library. Latencies are computed using the StarPerf constellation simulator on Kuiper satellites with ISL.

\vspace*{0.1cm}

\begin{figure}[t]
\vspace*{-0.2cm}
\centering
\includegraphics[width=0.495\textwidth]{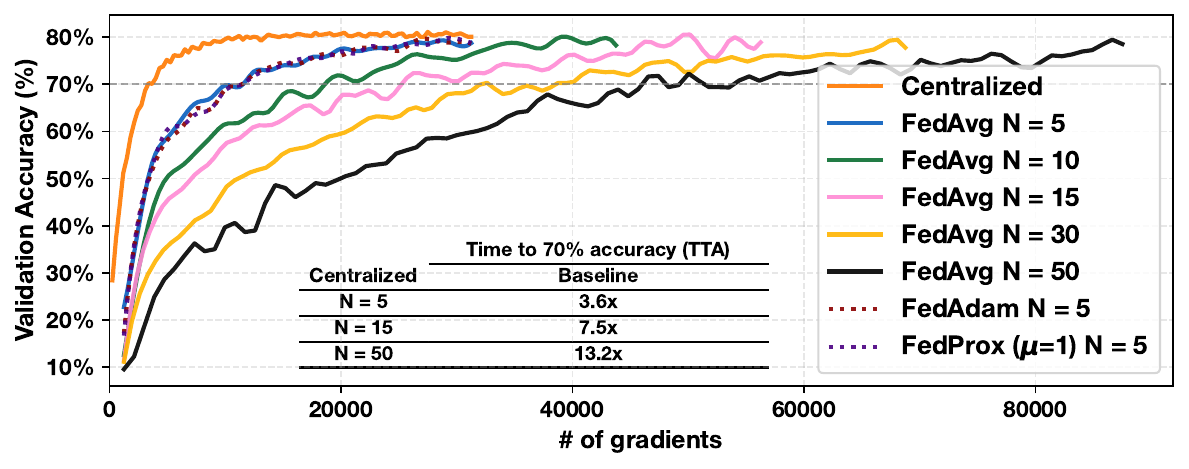}
\vspace*{-0.9cm}
\caption{Training accuracy: centralized vs. federated architectures}
\label{fig:training-results}
\vspace*{-0.6cm}
\end{figure}

\noindent\textbf{Training: Accuracy.}
 Figure \ref{fig:training-results} shows accuracy results 
 for centralized and different FL configurations. 
Centralized learning converges faster than FL. Time-to-accuracy for FL scales well with the number $N$ of satellites -- only X13 overhead for a X50 satellite increase for a centralized baseline. Aggregation optimizations have minimal impact on accuracy.

\noindent\textbf{Inference: Latency.}
 Table~\ref{tab:latency-table} shows inference results. 
Federated latency is highly scalable, independent from $N$. 
Centralized latency includes a hard physical lower-bound (RTT) and the GS model latency. How fast this last term  increases with $N$ depends on the parallelism level ($\alpha$) in the central server implementation. Scalability is thus more challenging. 
Overall, RTT in space induces a heavy penalty on the 
response time.

\begin{table}[h]
 \vspace*{-0.4cm}
\caption{\label{tab:latency-table}Inference latency results ($\downarrow$ is better)}
 \vspace*{-0.35cm}
\begin{centering}
\begin{tabular}{c|l|c}
\textbf{\# Sat.} & \textbf{Avg Lat. -- Centralized  (ms)} & \textbf{Avg Lat. -- Federated (ms)}\tabularnewline
\hline 
1 & 125.64 [124.20 + 1.44] & \multirow{3}{*}{23.75}\tabularnewline
\cline{1-2}
10 & 125.64 -- 134.28  & \tabularnewline
\cline{1-2}
100 & 138.60 -- 225 & \tabularnewline
\end{tabular}
\par\end{centering}
\vspace{0.05cm}
\scriptsize 
Cent. Avg. Lat. = RTT (124.2ms) 
+ $\alpha N\times$GS Inf. Lat. (1.44ms) with $\alpha=0.1$ to $0.7$
\scriptsize 
Testbed configuration: Desktop PC, CPU: AMD Ryzen 7 7800X3D, 32GB RAM, GPU: Nvidia RTX 4090, 24GB VRAM, WSL. 128 batch requests per satellite.\\
Centralized \& fed. inference lat.: PyTorch CUDA (GPU) \& CPU devices respectively. 

\scriptsize 

 \vspace*{-0.35cm}
\end{table}

\vspace*{0.05cm}

\noindent\textbf{Discussion and Next Steps}. 
Preliminary results show centralized designs tend to be the best option in the short term, security set aside, with faster training, but a hard physical lower bound on latency which requires distributed designs to go lower. 
In the longer term, decentralized designs might be preferable: training, while slower, yields much lower latencies for inference. However, such designs have clear benefits in terms of scalability at training and inference. 

Next, 
we intend to assess more in-depth security considering other threats such as  
tenant isolation~\cite{linFedSNFederatedLearning2025::short}. Digital twin~\cite{ferreiraReinforcementLearningSatellite2019::short} designs and split learning~\cite{sunEfficientPrivacyAwareSplit2024b::short} are promising to enhance security, respectively to place additional relevant counter-measures in the introduced mediation layer, and  
to avoid sending raw telemetric data by sharing inference between satellite and GS. 
Considering the imbalance between public vs. siloed data in terms of data volume, another promising avenue could be to explore decentralized learning approaches that share only a fraction of the model and yet enhance security~\cite{Shatter::short}.
Further experiments should also include more representative testbeds and datasets, with bigger models and HPC resources, closer to realistic large-scale space AI tasks.

\vspace*{-0.2cm}

\nocite{*}
\bibliographystyle{IEEEtran}
\bibliography{biblioshort}

\end{document}